\documentclass[submitting]{nst}

\usepackage{subfigure,dcolumn}
\usepackage[T2A,T1]{fontenc}
\usepackage[russian,english]{babel}

\newcommand{\xy}[1]{\textcolor{black}{#1}}
\newcommand{\nys}[1]{\textcolor{black}{#1}}

\newcommand{\gao}[1]{\textcolor{black}{#1}}
\usepackage{listings}

\let\textcite\relax
\begin{document}

\title{Development of a SciFi-based beam monitor for COMET}
\thanks{This work was supported in part by Fundamental Research Funds for the Central Universities (23xkjc017) in Sun Yat-sen University, the National Natural Science Foundation of China under Grant No. 12075326 and JSPS KAKENHI Grant No. 22H00139.}

% \author[a,1]{Yu Xu,}
\author{Yu Xu} 
\affiliation{School of Physics, Sun Yat-sen University, Guangzhou 510275, China}
\author{Yunsong Ning} \thanks{Yunsong Ning and Yu Xu contributed equally to the project.} 
\affiliation{School of Physics, Sun Yat-sen University, Guangzhou 510275, China}
\author{Zhizhen Qin}
\affiliation{State Key Laboratory of Particle Detection and Electronics, University of Science and Technology of China, Hefei 230026, China}
\affiliation{Department of Modern Physics, University of Science and Technology of China,  Hefei 230026, China}
\author{Yao Teng}
\affiliation{State Key Laboratory of Particle Detection and Electronics, University of Science and Technology of China, Hefei 230026, China}
\affiliation{Department of Modern Physics, University of Science and Technology of China,  Hefei 230026, China}
\author{Changqing Feng}
\affiliation{State Key Laboratory of Particle Detection and Electronics, University of Science and Technology of China, Hefei 230026, China}
\affiliation{Department of Modern Physics, University of Science and Technology of China,  Hefei 230026, China}
\author{Jian~Tang}\email[Corresponding author: tangjian5@mail.sysu.edu.cn, ]{}
\affiliation{School of Physics, Sun Yat-sen University, Guangzhou 510275, China}
\author{Yu~Chen,}
\affiliation{School of Physics, Sun Yat-sen University, Guangzhou 510275, China}
\author{Yoshinori~Fukao}
\affiliation{High Energy Accelerator Research Organization (KEK), Ibaraki 305-0801, Japan}
\author{Satoshi~Mihara}
\affiliation{High Energy Accelerator Research Organization (KEK), Ibaraki 305-0801, Japan}
\author{Kou Oishi}
\affiliation{Department of Physics, Imperial College London, London SW7 2AZ, UK}

\begin{abstract}
 COMET is a leading experiment to search for coherent conversion of $\mu^- \mathrm{N}\to e^- \mathrm{N}$ with a high-intensity pulsed muon beamline, produced by the innovative slow extraction techniques. Therefore, it is critical to measure the characteristics of the muon beam. We set up a Muon Beam Monitor (MBM), where scintillation fibers (SciFi) weaved in the cross shape are coupled to silicon photomultipliers (SiPM), to measure the spatial profile and timing structure of the extracted muon beam for COMET. The MBM detector has been tested successfully with a proton beamline in China Spallation Neutron Source (CSNS) and taken data with good performance in the commissioning run called COMET Phase-$\alpha$. Experience of the MBM development, such as the mechanical structure and electronics readout, and its beam measurement results will be shared.
\end{abstract}


\flushbottom
\maketitle

\section{Introduction}

With the steady increase of the beam power and intensity required for high precision measurement in particle and nuclear physics, a beam profile monitor plays an important role in precisely characterizing beam properties~\cite{He:2022jdo,Wang:2022evq,T2K:2014rve}, especially in the real-time control of the beam. The requirements for such a beam profile monitor include a fast response, high time resolution, and \gao{quasi-non-invasiveness} to the beam. 

Currently, there are several kinds of beam profile monitors.
\nys{The most widely used type of beam monitor is the Beam position monitors (BPMs) used to detect the position of the transverse beam, which can monitor the phase and transverse position of the beam in high-energy particle accelerators by measuring the difference in the total voltage between two opposite pick-ups~\cite{Li:2013bsa}. The BPMs have a wide variety of types, including button BPMs in the Chinese initiative Accelerator Driven Sub-critical system~\cite{Zhang:2016cee}, cavity BPMs for the International Linear Collider (ILC)~\cite{Walston:2007rj} and stripline BPMs in the KEKB injector linac~\cite{SUWADA2000307}.}
\nys{The other important category of beam monitor} is the gas detection system, which is typically based on the gas sheet~\cite{Yamada:2021dzs}. Due to its good stability and non-invasiveness to the beam, the gas detector is widely used for online beam profile measurement, such as the measurement of 400 MeV negative hydrogen atoms in J-PARC LINAC~\cite{Liu:2019qps,Kamiya:2018wbf,Miura:2015eca}, the muon beam measurement in NOvA experiments at Fermilab~\cite{Snopok:2019zhj}, and the electron/proton beam measurement in the High Luminosity LHC~\cite{Salehilashkajani:2021jav}.
\nys{Another} type of beam monitor is based on a multichannel plate (MCP)~\cite{Kim:2018aah,Sudjai:2022svy} and optical readout, which is applied in the muon g-2 experiment~\cite{Jegerlehner:2018zrj,Razuvaev:2017uty,Razuvaev:2019hmk} and planned to operate in CSNS~\cite{Zhang:2010kca} and HIRFL-CSR~\cite{Xie:2020xjp}. The working principle of MCP detector is that the injected electrons hit and avalanche amplified on MCP and then recorded by the anode collector board.  This design allows us to measure the beam profile for high-intensity beams, which can tolerate beam intensity up to  $10^{6} ~\mu/s$~\cite{Razuvaev:2017uty}. 

To achieve individual particle measurement, the scintillation fiber detector has become a good choice~\cite{Stoykov:2005yx} for its high light yield and fast response. Typically, the SiPMs are coupled to the scintillation fibers~\cite{Rossini:2022she} for photon detection. This type of beam profile monitor has been used in the R484/R582 experiments in RIKEN~\cite{Bonesini:2017nnl} and MEG-II experiment in PSI~\cite{DalMaso:2023hio}.
Furthermore, \nys{the scintillation fiber} detector can measure the hit time with high accuracy up to O(1) ns and the deposited energy by counting the photon electrons. Profit from the high-precision characteristics, it can also work for charged particle identifications through combination with multiple detection systems~\cite {Cohen:2016mdg}. 

The COMET  (COherent Muon to Electron Transition, J-PARC E21) experiment is a next-generation world-leading charged lepton flavor violation (cLFV) experiment, searching for the cLFV process via the coherent neutrinoless muon to electron conversion ($\mu-e$ conversion) process~\cite{COMET:2018auw}. 
COMET aims to measure the process with a single event sensitivity (S.E.S) of $2.7 \times 10^{-17}$, which is four orders of magnitude better than the current experimental limit given by SINDRUM-II~\cite{SINDRUMII:2006dvw}. \xy{An 8 GeV bunched proton beam with 1 MHz pulse structure is slowly extracted from the J-PARC main ring (MR)~\cite{Tomizawa:2022kqp,Oishi:2022doe,Noguchi:2022ioy}, which hits a stopping target and generates the required muon beam for the COMET experiment.} One of the key points in the experiment is the high-precision measurement of the COMET muon beam, which requires beam monitoring and measurement during the beam commissioning~\cite{Cerv:2014lfa}.

A muon beam monitor has been designed to measure and display the COMET muon beam. 
The goal is to offer a simple and cost-effective detector with a compact size and low power consumption, \nys{and the detector} must be easy to operate in the high radiation environment. This article will review the structure and performance of the detector and present the test results of the detector on high-intensity proton beams, and its performance during the COMET Phase-${\alpha}$ commissioning.

\section{Detector System}

\subsection{Mechanical structure}
The central of the MBM is a grid of multi-clad square 1-mm wide SCSF-3HF scintillation fibers produced by Kuraray~\cite{SciFi, 3HF} coupled at one end to S13360-1350PE SiPM from Hamamatsu~\cite{SiPM_cha}, 
forming a square beam window arranged along X and Y axes, where the orthogonal axis is perpendicular to the beamline direction. \nys{The effective photosensitive area of the selected SiPM is 1.3${\times}$1.3 mm${^2}$, which matches the cross-sectional area of the scintillation fiber, and the spectral response of the scintillation fiber and SiPM used is well matched, which can minimize the loss of photons at the transmission interface.} The use of square fibers makes the detector response independent from the position of the muon trajectory inside a fiber and minimizes the amount of dead space. The fibers are cut and polished by a diamond cutting tool, and the silicone grease is coated on the contact end to \nys{further} improve the photon transport efficiency via the fiber-SiPM interface. The fibers are divided into two perpendicular layers spaced 0.8 mm apart, which can cover a beam window of 30${\times}$30 cm${^2}$. Each layer is made of 128 fibers with a length of 500 mm and 1.3 mm spacing in order to measure the beam profile in the two orthogonal directions (X+Y).

\begin{figure}[htbp]
\centering
\includegraphics[width=0.9\hsize]{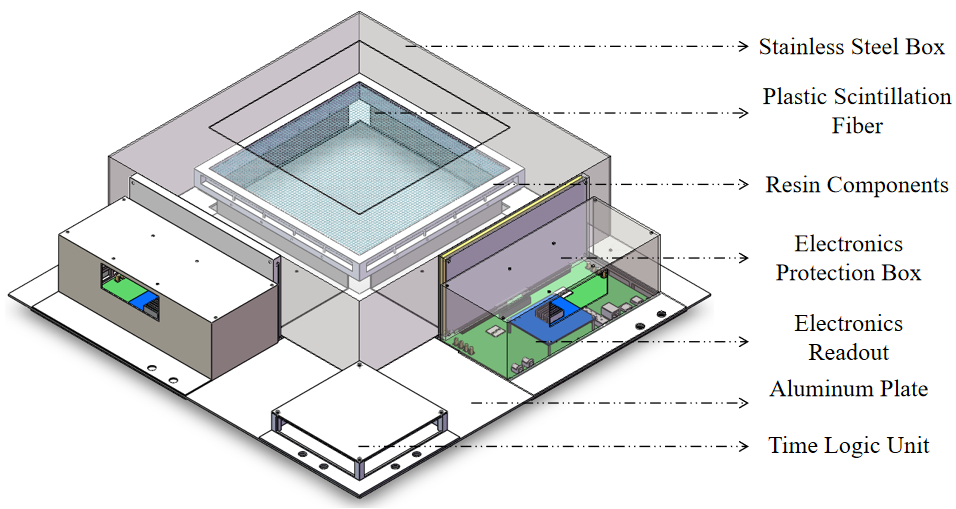}
\caption{The setup of the MBM structure} 
\label{MBM_Struct}
\end{figure}

\nys{Scintillation fiber has a total reflection cladding structure, which can achieve extremely high light collection efficiency and effectively improve the detection efficiency of the detector. When charged particles go through a scintillation fiber, energy loss occurs through ionization transfer in the scintillation fiber. The ionization energy loss can be estimated by the Bethe-Bloch formula~\cite{PDG}. For organic scintillator materials with a density close to 1 g/cm${^3}$, the corresponding minimum ionization energy loss dE/dx is about 200 keV/mm~\cite{Ruchti:1996ar}, which is acceptable for the high-intensity muon beam at the energy of $\mathcal{O}(10)$ MeV. At the same time, the mesh structure can further reduce the average energy loss of the muon beam so that such a detector structure can achieve quasi-non-invasive effects.}
By this quasi-non-invasive design and pure geometric calculation, we see that 60\% of the muons passing through the beam window will be recorded while 16\% of the muons will trigger both layers and make coincident signals. \nys{The signals will} provide the hit map of the muon and allow a characterization of the beam profile. 

All the \nys{scintillation fibers and their corresponding}  modules are fixed on the aluminum (Al) plate, and the grid of scintillating fibers is additionally shielded in a stainless steel box for the purpose of a light shield. The location of the beam window on the stainless steel box is cut off and covered with two layers of Al foil with a thickness of 50 $\mu$m. Additional to the grid of fibers shielded in the stainless steel box, we also install the electronic system on the Al plate, including two electronic modules and one Time Logic Unit (TLU). The two electronic modules are fixed on the two sides of the detector, with the SiPMs coupled to the end of the fibers. The TLU module is fixed at the corner of the Al plate \nys{for easy communication with the electronics on both sides. All electronic boards are covered by a resin shell with a metal layer on the surface, which can prevent dust and reduce the occasional single-particle flip damage caused by beam particle scattering to electronic components.} The whole scheme of the MBM is shown in Fig.~\ref{MBM_Struct}. 

\begin{figure}[!htb]
\centering
\subfigure[\label{Cosmic_test: a}]{
    \includegraphics[width=0.45\textwidth]{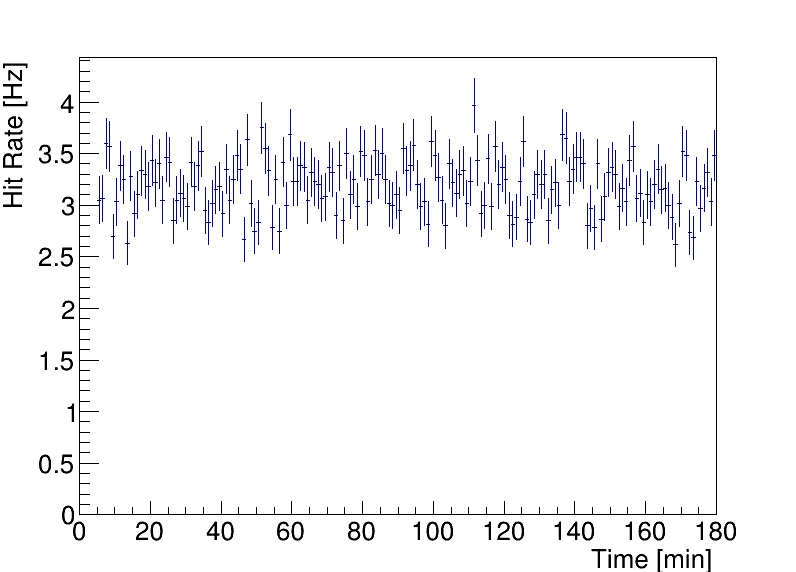}}
\quad
\subfigure[\label{Cosmic_test: b}]{
    \includegraphics[width=0.45\textwidth]{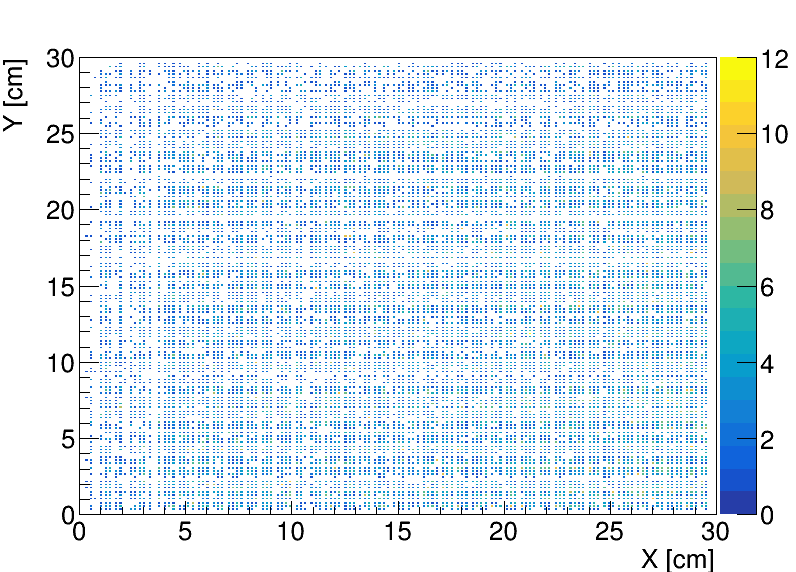}}
\caption{Tests with cosmic ray muons in the local laboratory. Panel (a) gives the trigger rate of cosmic ray muons by the MBM detector.  Panel (b) shows the hit map caused by cosmic ray muons. }
\label{Cosmic_test}
\end{figure}

\subsection{Electronics readout}
The electronics readout of the MBM includes two main parts: two electronic modules to read out the signals and upload data to the upper computer and one TLU to synchronize the clock time of the two electronic modules. 
Each electronic module contains three parts: SiPM Carrier, Frontend Electronics Board (FEB), and Data Acquisition (DAQ) Board.
The SiPM Carrier holds 128 pieces of SiPMs arranged in a 16${\times}$8 array and is connected to the FEB with a \nys{edge} connector. The SiPM will transfer the photons to photoelectrons and form a current signal, which will then be integrated by a charge-sensitive preamplifier, therefore a pulse signal with a fixed time length and a voltage of 3.3 V will be produced and processed in the FEB. A power-supply module is installed on the SiPM carrier to ensure a stable power supply to all 128 channels. The power supply module is equipped with a built-in temperature compensation system, which can adjust the voltage according to the external temperature, thereby improving the stability of SiPM in different environments. 
%\nys{In addition, the threshold of each SiPM can be set separately, which allows us to finely adjust the efficiency of different channels of the detector, ensuring that the detector remains in a uniform state as a whole.}

The FEB board mainly consists of two Application-Specific Integrated Circuit (ASIC) chips ~\cite{MaPMT_v10:2015} and one Field Programmable Gate Array (FPGA). The ASIC chip has 64 channels, each of which includes a charge-sensitive preamplifier, CR-RC shaping circuit, screener, and other modules. The ASIC chip, which is connected to the TLU in real-time through HDMI, will amplify the signals and transmit them to the FPGA. The TLU sends a clock and sampling start signal to the FEB, combining the signal with the timestamp generated by the timer inside the FEB, to achieve the time synchronization of SiPMs. The FPGA will pack the data and send them to the DAQ Board, which transfers data to the switch, and the data is stored in the computer (or MIDAS Bank) for further processing. \xy{We also made a graph interface software, which can receive data from MBM and draw the histogram for trigger rate in X and Y directions, to monitor the MBM running status in real-time.}

\xy{Due to the limit of the ASIC chip, the electronics system can only record the time information triggered by the signal while the charge information remains blank. Fortunately, counting and recording signals that pass through the threshold is sufficient to measure the beam profile and time structure. However, because the characteristics of SiPMs, such as the dark rate and quantum efficiency, are quite different, the response of the 256 channels of MBM would be also full of diversity, which would spoil the detector performance. In this case, we carefully design the electronics readout configuration so that it is possible to modify the threshold of each channel and adjust all channels in uniform responses to the signals. For each channel, we maintain the dark counting rate at a low level of approximately 1Hz by calibrating the threshold. In this way, we can expect a high signal-to-noise ratio in the beam monitoring run.}

\subsection{Validation of detector response}
After the assembly of the \nys{MBM}, \xy{it is necessary for us to validate the working status of the detector before installing it on the beamline. Thus, we continuously recorded cosmic ray data for around three hours.}
%and turning on all the electronic channels, we continuously took data for about 3 hours, in order to check the working status of the MBM, with the help of the cosmic ray muons in the local laboratory.
\xy{With this test, we are able to validate two questions: first, we cross check if the detector can run properly in the whole period; second, by comparing the collected data with our rough estimates, we check if the detector can record the charged particles (cosmic muons) with high efficiency.}

\xy{For the first purpose, we monitor the working status of the detector in the long run and find the stable data taking in the detector without any break, as shown with the trigger rate of this period in Fig.~\ref{Cosmic_test: a}.}

\xy{For the second purpose, we analyze the recorded data and compare them with our expectations. }The cosmic ray muon flux at the sea level is around 1 /min/cm$^2$~\cite{Bonechi:2019ckl}, so we must expect around 3 Hz in the MBM. In the actual configurations, by setting the appropriate threshold, the dark count rate is reduced to 1 Hz for each channel. Therefore, the trigger rate of all channels in the MBM remains high and stays at the level of O(100) Hz, which is about 2 orders of magnitude higher than the naive expectation from cosmic ray events. Fortunately, in this case, we can switch to the coincident measurement and collect the cosmic ray muon events, by restricting the X and Y layers triggered within 20 ns, which is 3$\sigma$ of the fiber decay time. \nys{Based on a coincidence time window of 20 ns, the accidental coincidence count rate caused by dark noise in two orthogonal scintillation fibers is 6.6 ${\times}$ 10${^{-4}}$ Hz, which is much lower than the muon count rate of 2.8 Hz based on Monte-Carlo simulation results. Therefore, this dark noise frequency is acceptable.}

Finally, we can get satisfactory results as shown in Fig.~\ref{Cosmic_test}. In Fig.~\ref{Cosmic_test: a}, we find the coincidence-triggered rate is about 3 Hz, which is consistent with our \nys{expected results}. Besides, the rate of the triggered events is stable in the plot, which indicates the detector's stability in the relatively long run. In addition, the detector response is almost uniform, which indicates a good performance for all the channels in MBM (see Fig.~\ref{Cosmic_test: b}). \nys{Based on the above observations, we believe that MBM can operate stably for a long time and has the requested functionalities to achieve physics objectives.}

\section{Beam tests}

\subsection{Proton beam test in CSNS}
To understand the detector response with a high-intensity beam, we \xy{conducted a quick beam test with the Associated Proton beam Experiment Platform (APEP) in the China Spallation Neutron Source (CSNS) on January 9th, 2023~\cite{Liu:2022czi}. The detector was installed at the platform on a movable bracket, which can move the detector on the X-Y plane with an accurate coordinate. Before the beam tests, we used a laser collimator to align the beam window to match the beam center.} 
%install the detector on the Associated Proton beam Experiment Platform (APEP) beamline in the China Spallation Neutron Source (CSNS) and conducted a 1-hour beam test on January 9th, 2023. 

\xy{For the setup of APEP, the extracted proton beam has a repetition rate of 25 Hz, }
%The proton beam has a repetition rate of 25 Hz, 
with a 400 ${\mu}$s \xy{beam pulse length} 
%spill time 
for each bunch. \xy{A suite of degraders is placed downstream of the proton beam window (PBW) and allows us to adjust the proton energy in a range of 10 to 80 MeV. In order to adjust the beam spot size and beam intensity, a collimation system is installed on the beamline. With this system, we are able to tune the proton uniform beam spot sizes at both the vacuum and air test points from 10 mm $\times$ 10 mm to 50 mm $\times$ 50 mm continuously. However, the scattering effect of atmospheric molecules on the beam spot in the non-vacuum experimental environment results in the actual beam spot size being significantly larger outside than its original size inside the vacuum beamline. For example, the setup of the collimation system of 10 mm${\times}$10 mm would result in the beam spot size of about 70 mm${\times}$70 mm at the air test point~\cite{Liu:2022czi}. } 

%\xy{We scanned 36 points uniformly distributed on the beam window area by moving the detector with the bracket, and each of the point was exposed in 90 seconds. The beam test results are listed as follows.} 
%The total beam density is 10${^8}$ p/s with 80 MeV/c momentum. We set the size of the collimator at 10 mm${\times}$10 mm, considering the scattering effect of atmospheric molecules on the beam spot in the non-vacuum experimental environment, the edge of the beam spot can spread to about 70 mm~\cite{Liu:2022czi}.
%The MBM was calibrated in position by a laser collimator and placed on a movable bracket. \nys{We uniformly selected 36 points on the detector beam window as beam detection points to check the performance of the entire area in the beam window.}%We scanned 36 points on the detector in order to check the performance of the whole area in the beam window.

\begin{figure}[htb]
\centering
\subfigure[\label{CSNS_Struct:a}]{
    \includegraphics[width=0.45\textwidth]{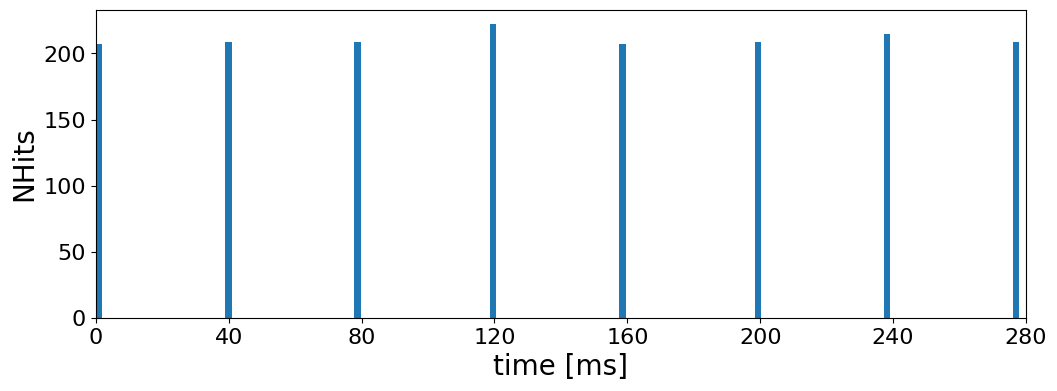}}
\quad
\subfigure[\label{CSNS_Struct:b}]{
    \includegraphics[width=0.45\textwidth]{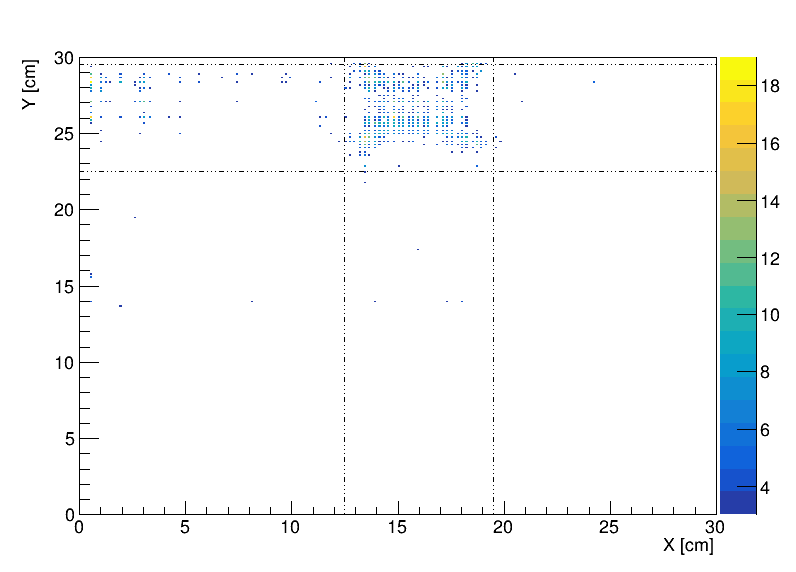}}
\caption{Beam test results with high intensity 80 MeV proton beam at CSNS. Panel (a) presents the timing structure of the proton beam in CSNS. Panel (b) shows the beam spot during CSNS beam tests. }
\label{CSNS_Struct}
\end{figure}

\xy{We have two purposes for the beam test. The first purpose is to validate the ability of the MBM detector to record the time structure of the beam.} 
%The first thing we want to check is if we can record the timing structure of the beam by MBM. 
\xy{Considering the 25 Hz repetition rate of the beam, we select the data within 280 ms, including 8 pulses. The time structure is shown in Fig.~\ref{CSNS_Struct:a}, and we see a clear time structure with a regular peak every 40 ms, which is consistent with the 25 Hz beam repetition rate.} 
%By showing the absolute trigger time of the events, we get the plot as shown in Fig.~\ref{CSNS_Struct:a}. We find a regular peak every 40 ms, which is consistent with the 25 Hz beam repetition rate. 

\xy{The other purpose is to validate the ability of the detector to measure the profile of the beam. Thus, we need to draw a 2-D plot by selecting the X-Y coincidence events.}
%The other thing we would like to check is if the detector can measure the beam profile. Therefore, we scanned 36 points uniformly distributed on the beam window area by moving the detector with the bracket. 
However, \xy{during the beam test,} we found a serious radiation effect for \nys{some} scintillation fibers operating in the 80 MeV proton beam \nys{for a long time}, in which almost all channels are triggered continuously at an ultra-high rate, especially for the previous scanned points. \xy{In this case, the true proton events are seriously polluted with the radiation backgrounds.} Thus, we have to make some stricter cuts to get a clean triggered signal:
\begin{itemize}
    \item The event should be triggered within 10 ns on the X and Y layers;
    \item The trigger time of the event should be in the range of the beam's spill time;
\end{itemize}
\xy{With these cuts, we get the result as shown in Fig. \ref{CSNS_Struct:b}. In the plot, we can see a clear beam spot in the top center, which is exactly the location of the beam arrangement. The size in the beam spot is about 70 mm${\times}$70 mm, which is also consistent with the simulation results given by CSNS~\cite{Liu:2022czi}. Based on these tests, finally, we can confirm that the MBM detector has a good response to the beam particles and can record the beam's timing structure and profile characteristics quite well.}
%With the cuts, we can get a cleaner spot, which is shown in Fig. \ref{CSNS_Struct:b}. In the plot, we can find a clear hot spot with a size of 70 mm${\times}$70 mm, which corresponds to the simulation results given by CSNS~\cite{Liu:2022czi}. The experimental results indicate that the MBM has a good response to the timing structure and profile characteristics of the beam.

\subsection{COMET Phase-${\alpha}$}
\xy{In order to measure the characteristics of proton beams and ${\pi}$/${\mu}$ production with less ambiguity, which is essential for the COMET experiment, the collaboration group proposed a low-intensity beam run called Phase-${\alpha}$ commissioning. One of the essential goals is to monitor and measure the properties of muon beams.} 
%The Phase-${\alpha}$ of COMET experiment aims to measure the characteristics of proton beams and ${\pi}$/${\mu}$ production with less ambiguity in a low beam intensity run. 
%It is extremely important to monitor and measure the properties of muon beams.
\xy{In total, }the detector system of Phase-${\alpha}$ consists of four \xy{sub detectors, }including MBM, Straw Tube Tracker, Range Counter (RC), and Proton Beam Monitor (PBM). As the first detector in the backend of the muon beam, MBM is used to monitor the timing structure and position information of the muon beam and provide a reference for the downstream detector and the beamline quality. \xy{The Straw Tube Tracker is installed after the MBM, and its task is to measure the position and direction of the injected particles~\cite{Nishiguchi:2022rcm}. Since the Straw Tube Tracker is part of the detection system in COMET Phase-I, the phase-$\alpha$ commissioning is to validate its performance in the beam environment, especially for its electronics system.}  
%The Straw Tube Tracker is part of the Phase-I detector and can measure the position and direction of the injected particles~\cite{Nishiguchi:2022rcm}. 
\xy{In the end is the Range Counter detector, which is made of} 
%The Range Counter is 
a series of several plastic scintillators coupled with photomultiplier tubes (PMT), aiming to measure the deposited energy and hit time of the muon events.
\xy{In addition, the electronics system of the Range Counter detector allows it to record about 10 $\mu$s PMT waveform, thus it has the ability to measure the decay time of muon decayed in orbit (DIO) events.} The PBM is installed on the proton beamline and provides a real-time monitor of the proton beam. \nys{Furthermore}, a dedicated beam-masking system is installed upstream of the transport solenoid entrance to study the optics and beam dynamics of the transport solenoid.

\subsubsection{Experiment setup}
\begin{figure}[htbp]
\centering
\subfigure[\label{Beam_Struct:a}]{
    \includegraphics[width=0.45\textwidth]{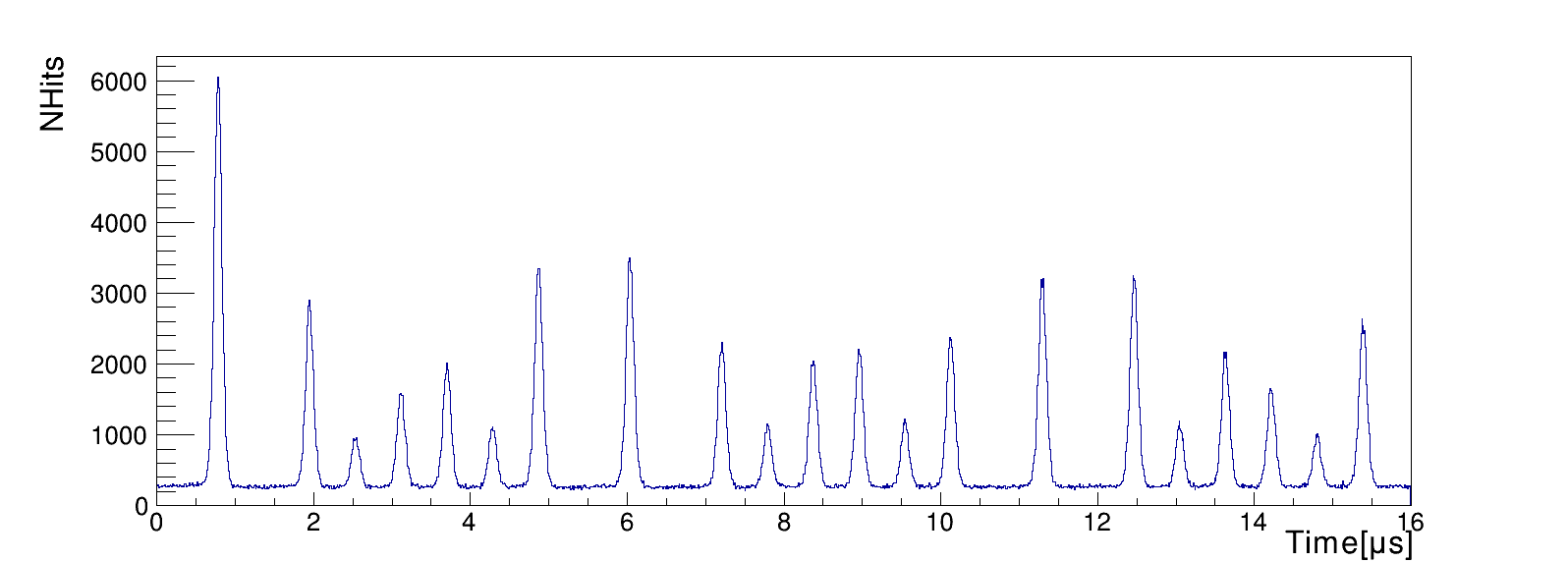}}
\quad
\subfigure[\label{Beam_Struct:b}]{
    \includegraphics[width=0.45\textwidth]{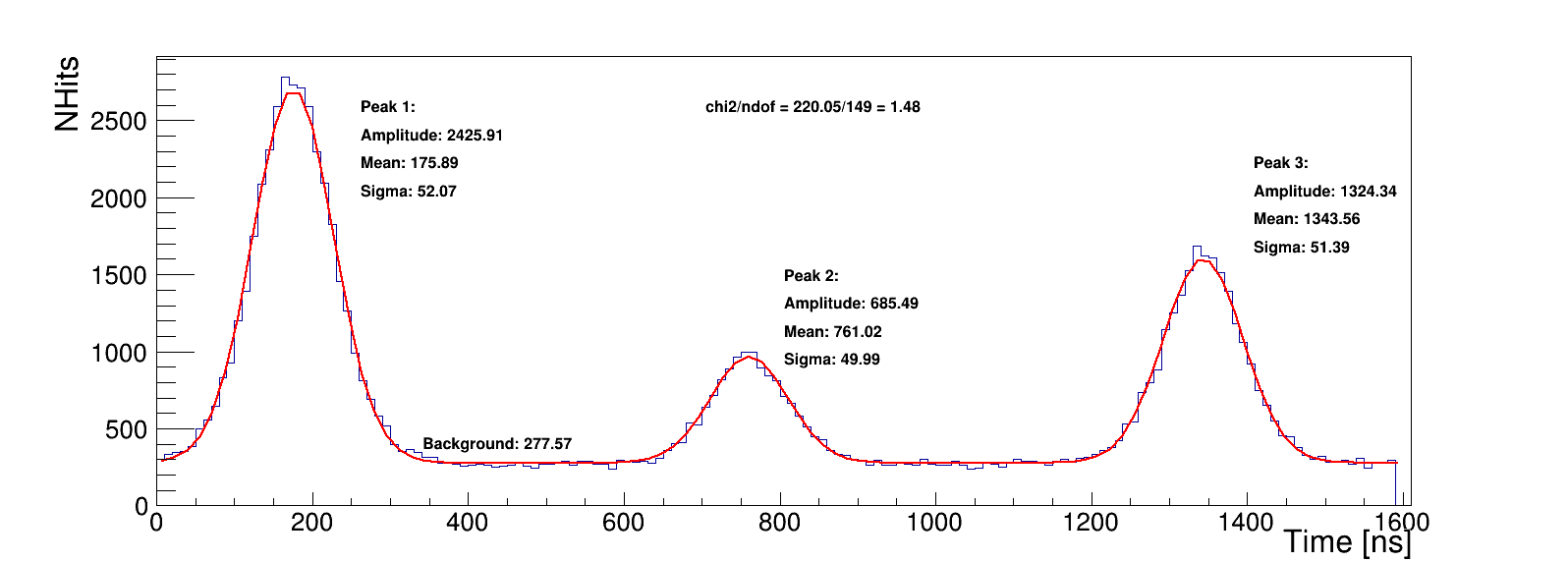}}
\caption{Monitor the timing structure of the muon beam in the phase-$\alpha$. Panel (a) shows the overview of the timing structure; Panel (b) is the zoom-in details and peak fittings in the short time window.}
\label{Beam_Struct}
\end{figure}

The detector was installed in the COMET Experimental Hall in the middle of February 2023. A \xy{dedicated} 8 GeV \xy{bunched} proton beam is set up at the J-PARC Main Ring (MR), which hits the graphite target and produces the secondary muon beam. \nys{This energy proton beam can minimize antiproton pollution and generate enough muons to meet physical requirements.~\cite{COMET:2018auw}} The beam spill cycle \xy{time} is \xy{set to} 9.2 s, \xy{in which the acceleration time and the flat top time is} about 0.6 s \xy{and 8.6 s respectively}. \xy{The beam power in the Phase-$\alpha$ commissioning is 0.26 kW.} \nys{The original plan is to use a 3 T magnetic field to guide the muon beam inside the transmission solenoid. However, in reality, the magnetic field used was only 1.5 T, resulting in a wider extension of muon beams.}
%A 1.5 T magnetic field was set to guide the muon beam inside the transport solenoid, which is half of our initial plan and leads to a much wider spread of the muon beam. 
The MBM was installed just after the exit of the transport solenoid, with its center aligned to the center of the transport solenoid exit. In addition, during installation, we introduced a trigger signal called 'local time' provided by the Range Counter through an SMA interface, which is used to reset the time corresponding to MBM to zero when the event is triggered. We took data from March 3rd to 5th and 9th to 15th with the secondary muon beam.

\subsubsection{Experiment Results}
\xy{For the Phase-$\alpha$ commissioning, the goal of MBM includes two points: first, we need to characterize the time structure of the muon beam and compare it with the bunched proton time structure from the J-PARC Main Ring; second, we would like to measure the muon beam spatial profile, to have a deeper understanding of the beam production and transportation. Moreover, it is always wise to assess the operational stability of MBM under a harsh environment in the long run.}

The timing structure of the muon beam in the COMET is one of the key questions we are concerned about. \xy{As we mentioned before, COMET needs a dedicated 8 GeV bunched proton beam which generates the secondary muon beam. In the design of the COMET experiment, the J-PARC Main Ring synchrotron accelerates the proton cycle by cycle, and one Main Ring cycle consists of 3 bunches (type a) spaced by 1.17 $\mu$s and one bunch spaced by 1.76 $\mu$s (type b)\nys{, which is believed to effectively reduce the physical background caused by other beam particles}. Owing to the low beam intensity in the Phase-$\alpha$ commissioning, there are on average only 0.02 events recorded by MBM in one bunch, thus we have to superpose many events to get the plot with a visible structure. However, due to the limit of the timing resolution from the Main Ring, we are not able to confirm the precise number of the bunch for the triggered hit. Thus, the time structure should be a superposition of four possibilities (aaab, aaba, abaa, baaa). In this way, we expect a time structure of one big peak and six small peaks, while the big peak must be spaced in 1.1 $\mu$s compared to small peaks, and the small peaks must be separated by 586 ns in the zoomed time window.} By accumulating data in around 800 runs, we managed to draw the timing structure of the muon beam, as shown in Fig.~\ref{Beam_Struct}, \xy{which meets our expectations very well}. 
%From the accelerator side, we know that the accelerator main ring cycle consists of 3 peaks separated by 1.17 $\mu$s, then follows a peak 1.76 $\mu$s later.
%Since we get the plot by accumulating many runs, and the start of the run can be any of the four peaks, thus we should expect one main peak followed by six smaller peaks, while the time distance from the main peak and the small peaks are 1.1 $\mu$s, and the time distance between the six small peaks is 586 ns, which matches what we observe from the figure very well. 

\begin{figure}[!htb]
\centering
\includegraphics[width=0.45\textwidth]{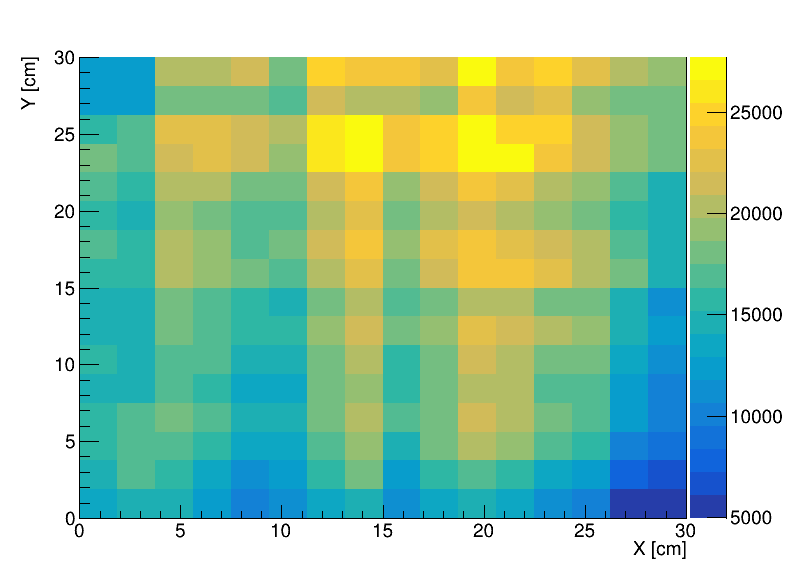}
\caption{The 2D muon beam profile in COMET Phase-$\alpha$.} 
\label{COMET_hitmap}
\end{figure}

The other important thing we are concerned about is the muon beam profile of COMET. We managed to get the 2D profile of the muon beam, as shown in Fig~\ref{COMET_hitmap}. \xy{In the actual beam trial operation stage, the measured beam profile is not completely circular result of interference from many factors. One of the reasons is the magnetic field in the transport solenoid set up to 1.5 T, which is half of what was originally designed in the COMET Phase-I design. Besides, the detectors in Phase-${\alpha}$ are installed out of the exit of the transport solenoid without any magnetic field, let alone the scattering effects in the air. In addition, the beam must contain several kinds of charged particles, including e$^-$, $\mu^-$, and $\pi^-$, each of which has a different beam direction. Based on the above points analysis, the beam is supposed to be widely spread. This must result in different beam spots in the beam profile.}
%However, because of the wide spread of the muon beam, we cannot get an obvious full beam spot with the MBM. 
\begin{figure}[!t]
\centering
\subfigure[\label{Run_Stable:a}]{
    \includegraphics[width=0.45\textwidth]{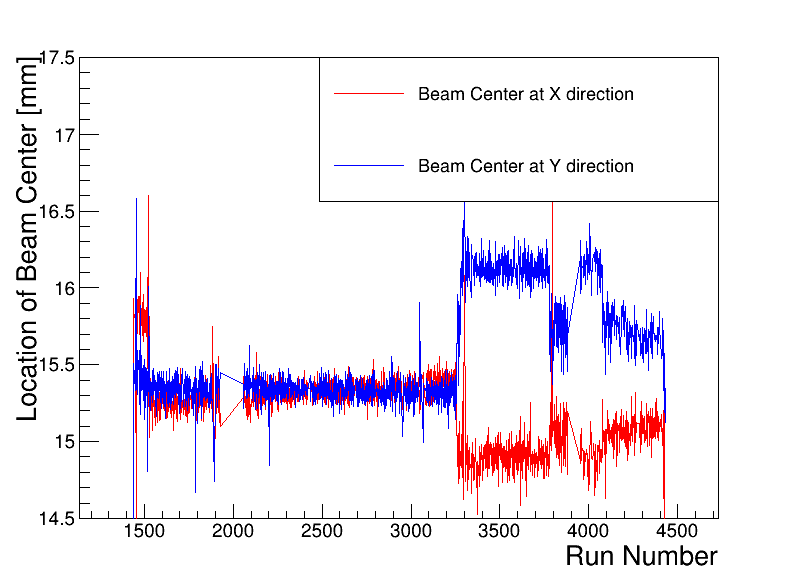}}
\quad
\subfigure[\label{Run_Stable:b}]{
    \includegraphics[width=0.45\textwidth]{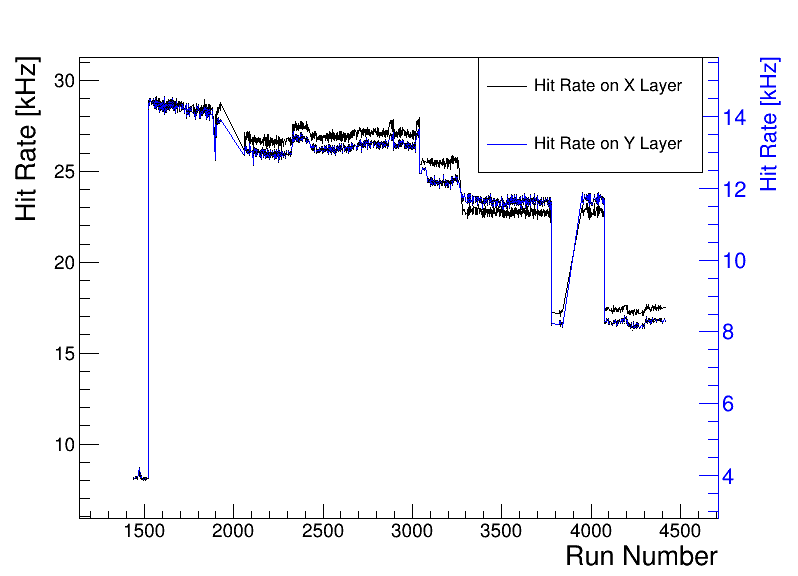}}
\caption{The long-term stability of the muon beam. Panel (a) shows the change of center location during the beam time; Panel (b) monitors the change of hit rates during the beam time.} 
\label{Run_Stable}
\end{figure}

\xy{The Phase-${\alpha}$ commissioning start from March 3rd to March 13th with 1.5 T magnetic field.} 
\nys{During this period, the muon beam was operated stably, and we conducted detailed measurements of the time structure and the beam profile structure. The results can be reflected by the stable beam spot in Fig.~\ref{Run_Stable}.}
In order to obtain more comprehensive beam profile features, 
\xy{we draw the barycenter of the beam profile} and the counting rate of the detector, \xy{which also indicates the long-term stability of the detector during the whole Phase-${\alpha}$ commissioning.}
%The platform period of the counting rate in the picture indicates the long-term stability of the detector during Phase-${\alpha}$ commissioning. 

\xy{After the successful run in the muon mode, we turned off the magnetic field for several hours and then inverted the direction of the magnetic field and measured the anti-muon component in the beam, which is called "Mu+ Run", in order to study the beam component and the effect of the magnetic field applied to the beam. In addition, we moved the location of the beam masking system several times during the "Mu+ Run" period, to validate the beam optics and dynamics in the curved transport solenoid. Thus, in Fig.~\ref{Run_Stable}, we also see the location of the beam center and the stable trigger rate most of the time. It is noted that the beam jumps several times after Run 3257, which is caused by the swap of the magnetic field polarization and the movement of the beam masking system. Generally speaking, the COMET muon beam runs stably and the MBM can reflect the running status of the beam quite well.}

\begin{figure}[htbp]
\centering
\subfigure[\label{Coin_RC:a}]{
    \includegraphics[width=0.45\textwidth]{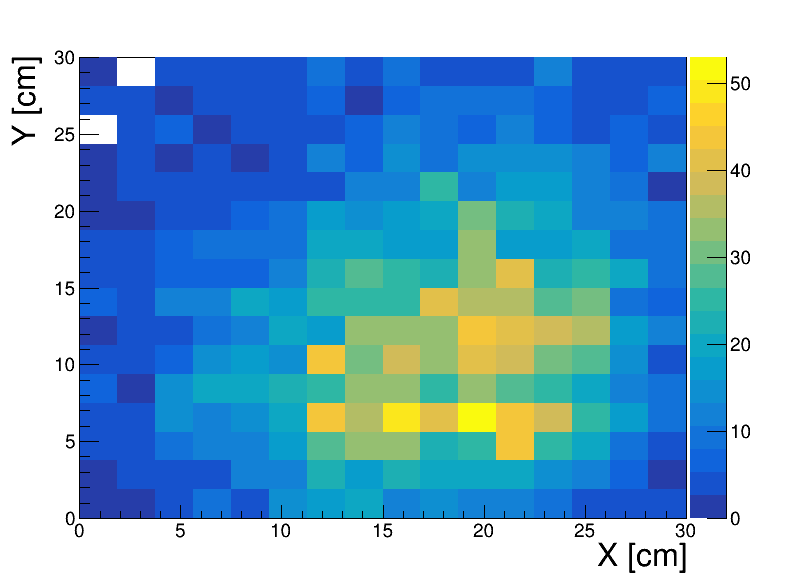}}
\quad
\subfigure[\label{Coin_RC:b}]{
    \includegraphics[width=0.45\textwidth]{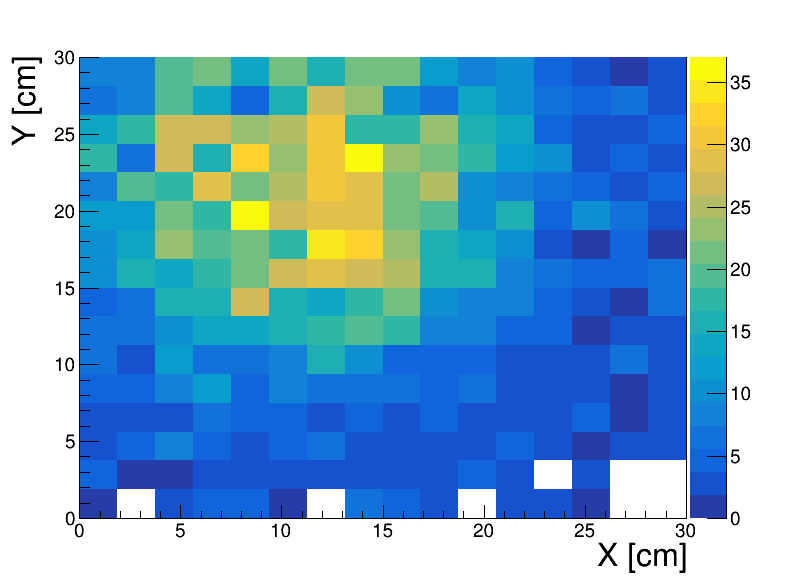}}
\caption{The MBM-RC coincidence hit map and the movement of RC during the commissioning. Panel (a) shows that RC is located in the bottom right while Panel (b) displays that RC is moved to the top left.} 
\label{Coin_RC}
\end{figure}

\nys{To cross-check whether the detector can correctly identify the beam position}, we also conduct the coincidence measurement with Range Counter by selecting the first hits in every event. As a result, we get the hit maps shown in Fig.~\ref{Coin_RC}. We see a clear hot zone in the plots, which indicates the position of the Range Counter. 
%\xy{During March 12th to 13th,} 
We moved the position of the Range Counter \xy{to several different locations, which allowed us to validate the MBM's response to the beam location. We record the relative location of the Range Counter in this period and draw the MBM-RC coincidence beam profile with the corresponding data. In Fig.~\ref{Coin_RC}, we can see the beam spot located at the bottom right (Fig.~\ref{Coin_RC:a}) and the top left (Fig.~\ref{Coin_RC:b}), which corresponds to RC movement from the bottom right to the top left. With these results, we verify the correctness of the MBM response to the beam.} 
%during the experiment, which is consistent with the coincident measurement result that the center of the beam spot moved from the bottom right (Fig.~\ref{Coin_RC:a}) to the top left (Fig.~\ref{Coin_RC:b}).

\section{Summary and outlook}
%\section{Conclusion}

The first complete Muon Beam Monitor with 1 mm wide scintillation fibers and X-Y readout by SiPM has been designed and used for COMET Phase-${\alpha}$ commissioning. The detector has reached a time resolution of 1.6 ns and a spatial resolution of 1 mm, with a beam window of 30${\times}$30 cm${^2}$. We have validated the \nys{monitor} system with cosmic ray muons in the local laboratory and checked the MBM functionalities with several rounds of beam tests, including the timing structure and beam profile of the muon beam and proton beam. %\nys{The performance of MBM has performed well in all beam tests, demonstrating its ability to be applied to various beams and having broad usage space.}
The performance of MBM is in a good shape. The current experience in the development of MBM for the COMET experiment might facilitate other similar beam instrumentations in various accelerator centers. 

From the experience of the \xy{COMET} Phase-${\alpha}$ commissioning, we are considering a further upgrade on the MBM to satisfy the requirements for the operation in the Phase-I period. \nys{Compared to the current Phase-${\alpha}$, the beam intensity in Phase-I will be increased by about three orders of magnitude. It is expected to upgrade the electronic readout of the detector to avoid pile-up events, including an upgrade of electronics to record both charge and time information, and a more powerful DAQ system for the much higher trigger rates. Moreover, as the backend detector needs to be installed inside the Detector Solenoid (DS) in the Phase-I experiment, and the DS will provide a 1 T high magnetic field environment~\cite{Sumi:2023wfk}, the electronic parts to collect SiPM signals need to be placed outside the DS to avoid damage caused by high magnetic field environment. However, the location of electronic components also requires a consideration of the strong radiation effect caused by neutrons in the harsh environment outside the solenoid. Therefore, it will be a tough task to strengthen the radiation hardness of the electronics to avoid single-event flipping. There is always a long way to go to pin down the exclusion limit or claim the discovery of new physics. Nevertheless, this deserves an effort to make a breakthrough and is the way to drive cutting-edge technology with fundamental science.}
%the vacuum interface for MBM parts also need to be considered.
%\nys{In addition to improving the electronic triggering rate to adapt to higher intensity beam currents, it is also necessary to improve the mechanical structure. At present, the position accuracy of the detector depends on the cross-section of the scintillating fiber. In order to further improve the position resolution of the detector, the mode of the fiber strip provides a good reference. The combination of fiber optic strips and SiPM arrays makes it possible to use the center of gravity reconstruction method in measurement, which can effectively improve accuracy and make measurement accuracy higher than physical accuracy. In addition, due to the use of higher intensity magnetic fields, the beam size will be reduced due to the enhanced beam collection ability, so the required detector beam window area can be reduced to simplify the installation plan. More importantly, the current signal output by the detector is a digital signal that only provides quantity and time, lacking amplitude and energy information, making it impossible to distinguish particle signals. Therefore, it is necessary to perform more detailed processing on the simulated signal in order to obtain the energy information of the particles for further beam analysis in the future.}

\section*{Acknowledgments}
We thank the J-PARC as the host laboratory and the COMET phase-$\alpha$ team for their strong support. We are grateful to the CSNS accelerator group for offering the stable beam during, two rounds of tests. We would also like to thank Prof. Yi Liu from Zhengzhou University for his contribution to DAQ.

% \newpage
%\printbibliography

\end{document}